\def\etal{{\it et al.\thinspace}}
\def\mearth{{\rm\,M_\oplus}}

\documentclass[apjl]{emulateapj}

\begin{document}

\shorttitle{The Search for other Earths}
\shortauthors{S.~N.~Raymond}

\title{The Search for other Earths: limits on the giant planet 
orbits that allow habitable terrestrial planets to form}

\author{Sean N. Raymond\altaffilmark{1,*}}

\altaffiltext{1}{Laboratory for Atmospheric and Space Physics,
University of Colorado, UCB 392, Boulder CO 80309-0392;
raymond@lasp.colorado.edu}
\altaffiltext{*}{Member of NASA Astrobiology Institute}

\begin{abstract}

Gas giant planets are far easier than terrestrial planets to detect around
other stars, and are thought to form much more quickly than terrestrial
planets.  Thus, in systems with giant planets, the late stages of terrestrial
planet formation are strongly affected by the giant planets' dynamical
presence.  Observations of giant planet orbits may therefore constrain the
systems that can harbor potentially habitable, Earth-like planets.  We present
results of 460 N-body simulations of terrestrial accretion from a disk of
Moon- to Mars-sized planetary embryos.  We systematically vary the orbital
semimajor axis of a Jupiter-mass giant planet between 1.6 and 6 AU, and
eccentricity between 0 and 0.4.  We find that for Sun-like stars, giant
planets inside roughly 2.5 AU inhibit the growth of 0.3 Earth-mass planets in
the habitable zone.  If planets accrete water from volatile-rich embryos past
2-2.5 AU, then water-rich habitable planets can only form in systems with
giant planets beyond 3.5 AU.  Giant planets with significant orbital
eccentricities inhibit both accretion and water delivery.  The majority of the
current sample of extra-solar giant planets appears unlikely to form habitable
planets.

\end{abstract}

\keywords{planetary systems: formation ---
methods: n-body simulations --- astrobiology}

\section{Introduction}

In systems with gas giant planets, there exists an unavoidable link between
terrestrial and giant planets.  Giant planets are constrained to form during
the few million year lifetime of the gaseous component of protoplanetary disks
(Haisch \etal 2001).  On the other hand, terrestrial planets take tens of
millions of years to form in a bottom-up fashion from km-sized planetesimals
through Moon- to Mars-sized planetary embryos (see review by Chambers 2004).
Thus, giant planets are present during the late stages of terrestrial
accretion.  It is during this late stage that water may be accreted by
terrestrial planets in the form of water-rich bodies originating past $\sim$
2.5 AU (Morbidelli \etal 2000; Raymond, Quinn \& Lunine 2004 -- hereafter
RQL04; see Drake \& Righter 2002 for a different opinion).  Indeed, the
gravitational influence of the giant planets can shape the orbital and
compositional characteristics of systems of terrestrial planets (e.g.,
Chambers \& Cassen 2002; Levison \& Agnor 2003; RQL04).

Estimates of the fraction of Sun-like stars with massive planets range from
5\% to 25\%, although these values depend on the semimajor axes of the planets
in question (Tabachnick \& Tremaine 2002, Lineweaver \& Grether 2004, Fischer
\& Valenti 2005).  This fraction appears to be significantly lower for
low-mass stars (Endl \etal 2006).  In addition, Greaves \etal (2006) saw no
correlation between the presence of debris disks and giant planets, suggesting
that many terrestrial planet systems may not contain giant planets.  By
studying the link between giant and terrestrial planets, we are restricting
ourselves to the subset of planetary systems containing giant planets; note
that this subset still comprises billions of stars in our Galaxy.

Previous work has documented correlations between the giant and terrestrial
planets.  Systems with more massive giant planets tend to form fewer, more
massive terrestrial planets than systems with less massive giant planets
(Levison \& Agnor 2003; RQL04).  In addition, giant planets with significant
orbital eccentricities preferentially eject water-rich material in the outer
disk and cause the terrestrial planets to be dry (Chambers \& Cassen 2002;
RQL04).

Several studies have used massless test particles to search for stable regions
in the known extra-solar systems of giant planets (e.g., Jones \etal 2000;
Rivera \& Lissauer 2001; Menou \& Tabachnik 2003; Barnes \& Raymond 2004).
These stable regions are considered locations where Earth-like planets might
exist.  However, this method is limited because forming planets do not behave
as massless bodies -- indeed, such studies would predict the existence of an
Earth-sized planet at 3 AU in the asteroid belt.  However, gravitational
jostling among embryos in moderately stable regions like the asteroid belt
(which is pervaded by resonances with the giant planets) removes the vast
majority of the mass in the region (Wetherill 1992).  For example, Barnes \&
Raymond (2004) found stable regions for test particles in four extrasolar
planetary systems, but later work showed that terrestrial planets could only
form in two of the systems (Raymond, Barnes \& Kaib 2006).  Only a few authors
have studied the formation of terrestrial planets in known extra-solar systems
(Th\'ebault \etal 2002, 2004; Quintana \etal 2002; Raymond, Barnes \& Kaib
2006).

In this {\it Letter} we derive limits on the extra-solar giant planet systems
that are likely to harbor potentially habitable planets.  This is designed as
a guideline for upcoming missions searching for extra-solar Earths, such as
ESA's {\it Darwin} and NASA's {\it Kepler} and {\it Terrestrial Planet
Finders}.  Specifically, our goal is to determine the region in the orbital
parameter space of a giant planet where potentially habitable planets can form
in the habitable zone.

With a sample size of one, it is difficult to decide which theoretical planets
could harbor life.  For this analysis, we require potentially habitable
planets to 1) form in the circumstellar habitable zone (Kasting \etal 1993),
2) have a minimum mass of 0.3 Earth masses ($\mearth$) -- below this value
planets are unlikely to sustain plate tectonics for many Gyr (Williams \etal
1997; Raymond, Scalo \& Meadows 2006), and 3) have a significant water
content.  Not only is water thought to be vital for life, it may also help
with plate tectonics (Regenauer-Lieb \etal 2001).

We simulate the formation of terrestrial planets from a disk of planetary
embryos in the presence of a single, Jupiter-mass giant planet.  We
systematically vary the giant planet's orbital semimajor axis between 1.6 and
6 AU, and its eccentricity between 0 and 0.4.  Simulation outcomes constrain
the systems in which terrestrial planets can form, and provide an
observational test of which extra-solar systems are able to form habitable
planets.  We extend our analysis to stars of different masses and compare with
the known sample of extra-solar giant planets.

\section{Simulations}

We start from a protoplanetary disk that extends from 0.5 to 4 AU with a
surface density distribution that scales with radial distance $r$ as
$\Sigma(r) = \Sigma_1 r^{-3/2}$.  We choose $\Sigma_1$, the surface density at
1 AU, to be 10$g\,cm^{-2}$, about 50\% higher than the minimum-mass solar
nebula model (Hayashi 1981).  We assume that planetary embryos have formed in
the disk, spaced by $\Delta = $5-10 mutual Hill radii, as predicted by models
of oligarchic growth (e.g., Kokubo \& Ida 1998).  We generate four disks, each
with 46-49 embryos.  Embryo masses $M$ increase as $M \propto \Delta^{3/2}
r^{3/4}$ (e.g., RQL04) and range from roughly 0.03 to 0.3 $\mearth$, totalling
6.2 $\mearth$.

Embryos start with a water distribution that reflects the current distribution
of water in primitive classes of asteroids (Abe \etal 2000; see Fig. 2 from
RQL04): inside 2 AU embryos are dry, outside 2.5 AU they are wet (5\% water by
mass) and from 2-2.5 AU they contain a moderate amount of water (0.1\% water
by mass; see first panel of Fig.~\ref{fig:aet}).  We also assign embryos a
starting iron content that is interpolated between the known values for the
planets (as in Raymond \etal 2005a, 2005b), including a dummy value of 40\%
iron by mass in place of Mercury because of its anomalously large iron
content.

We assume that one Jupiter-mass giant planet exists in the system.  We vary
its semimajor axis $a_J$ between 1.6 and 6.0 AU (spaced every 0.2 AU), and its
eccentricity $e_J$ between 0.0 and 0.4 (spaced every 0.1).  These initial
conditions inherently assume that embryos form more quickly than giant planets
(as in the core-accretion scenario; e.g., Pollack \etal 1996).  In cases with
small $a_J$ or large $e_J$, embryo formation may be unlikely in certain
regions of the disk (due to strong resonances and high planetesimal
velocities).  However, those embryos whose formation is in question are
quickly removed from the system via dynamical ejection.  Thus, we consider our
simulations a reasonable approximation of terrestrial accretion even if giant
planets form more quickly than embryos (e.g., Boss 1997).

For each ($a_J$,$e_J$) combination we perform four simulations of terrestrial
planet growth, i.e. one for each disk of embryos we've generated, for a total
of 460 simulations.  We integrate each simulation for 200 million years with a
6 day timestep using the hybrid integrator {\tt Mercury} (Chambers 1999).
Collisions are treated as inelastic mergers conserving water and iron (for a
discussion, see Raymond, Quinn \& Lunine 2006b).  Each simulation conserved
energy to better than one part in $10^{3}$.

Low-resolution simulations such as these can reproduce the bulk properties of
the Solar System (Agnor \etal 1999, Chambers 2001).  So, although a given
simulation may lack certain details of a high-resolution simulation (Raymond,
Quinn \& Lunine 2006a), the large number of simulations we run should
reproduce the basic properties of the planetary systems.  Indeed, since
computational speed scales with the number of particles $N$ as $N^2$,
low-resolution simulations are ideal for explorations of parameter space.

\medskip
\epsfxsize=8truecm \epsscale{0.5} \epsfbox{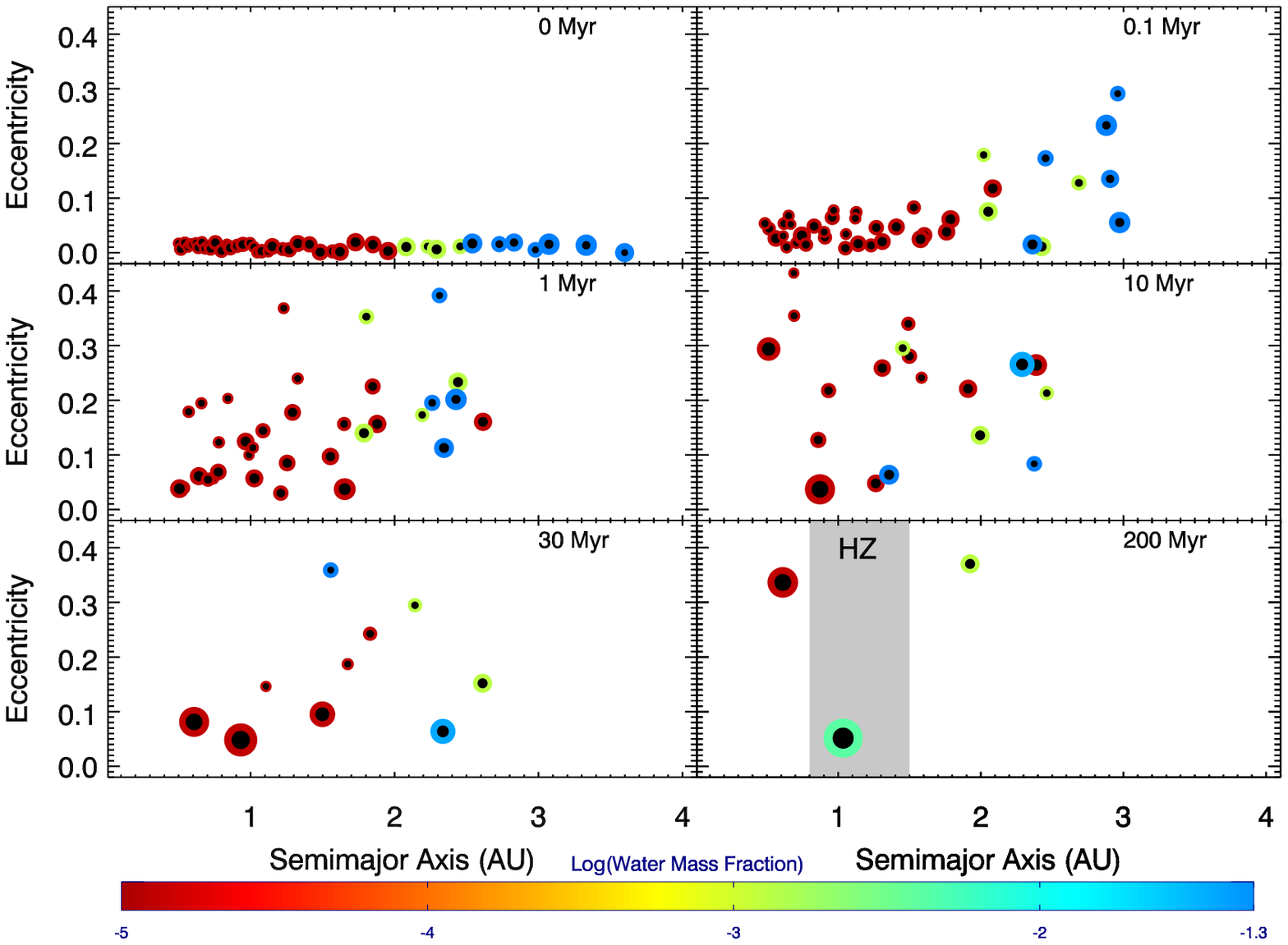}
\figcaption[f1.eps]{\label{fig:aet} \small{Snapshots in time of the evolution
of a simulation with a zero eccentricity giant planet at 4.2 AU.  The
size of each body indicates its relative physical size, but is not to
scale.  The color represents its water content, and the dark circle in
the center shows the size of its iron core.  In this case a
water-rich, 1.6 $\mearth$ planet formed at 1.03 AU in the HZ (shaded
in the last panel).}}
\medskip

Figure~\ref{fig:aet} shows the evolution of a simulation with ($a_J$,$e_J$) =
(4.2 AU, 0.0), color-coded by water content.  The outer regions of the disk
are excited by interactions with Jupiter (e.g., the 2:1 mean motion resonance
at 2.6 AU), and the inner disk is excited by gravitational perturbations among
embryos.  As eccentricities increase, orbits cross and collisions occur.  In
time, planets grow and the number of bodies dwindles.  Many embryos are
scattered from their original locations, sometimes delivering water-rich
material to planets in the inner regions.  Water delivery occurs relatively
late in the evolution, because multiple scattering events are needed for
significant radial movement (Raymond, Quinn \& Lunine 2006a).  In this case
three terrestrial planets formed, at 0.61, 1.03 and 1.94 AU.  The planets at
1.03 and 1.94 AU accreted material from past 2.5 AU, but the inner planet is
dry.

Figure~\ref{fig:aem} shows the regions in ($a_J$,$e_J$) space where
potentially habitable terrestrial planets can form.  Crosses mark regions
where no terrestrial bodies survive in the habitable zone (HZ), defined to lie
between 0.8 and 1.5 AU (see shaded region in '200 Myr' panel from
Fig.~\ref{fig:aet}).  Red dots indicate regions where terrestrial bodies
survive in the HZ but average less than 0.3 $\mearth$.  Green dots mark where
the average mass of terrestrial planets in the HZ exceeds 0.3 $\mearth$, but
these planets have not accreted water-rich material from past 2-2.5 AU.  Blue
dots indicate where planets in the HZ have average masses greater than 0.3
$\mearth$ and average water contents greater than $5\times10^{-4}$ by mass (a
rough lower limit on the Earth's water content).

Figure~\ref{fig:aem} shows clear limits to where habitable planets can form,
despite some scatter due to the stochastic nature of the accretion process
(e.g., the lack of 0.3 $\mearth$ planets for ($a_J$,$e_J$) = (5.2 AU, 0.1) and
(6.0 AU, 0.0) and their presence at (1.8 AU, 0.0)).  Habitable-mass ($>0.3
\mearth$) planets can form in the HZ if $a_J >$2.5 AU.  Water-rich
habitable-mass planets can only form for $a_J >$ 3.5 AU.  The criterion that
planets must accrete water-rich embryos (i.e., the blue dots) is ``noisier''
than our 0.3 $\mearth$ planetary mass criterion because the spacing between
embryos increases with orbital distance, such that only $\sim$ 10 water-rich
embryos exist at the start of each simulation.  Small-number statistics is
therefore responsible for the noise in the data.  However, high-resolution
simulations have demonstrated that the water delivery process is actually much
more robust (less stochastic) than previously thought (Raymond, Quinn \&
Lunine 2006b).

\medskip
\epsfxsize=8truecm \epsscale{0.5} \epsfbox{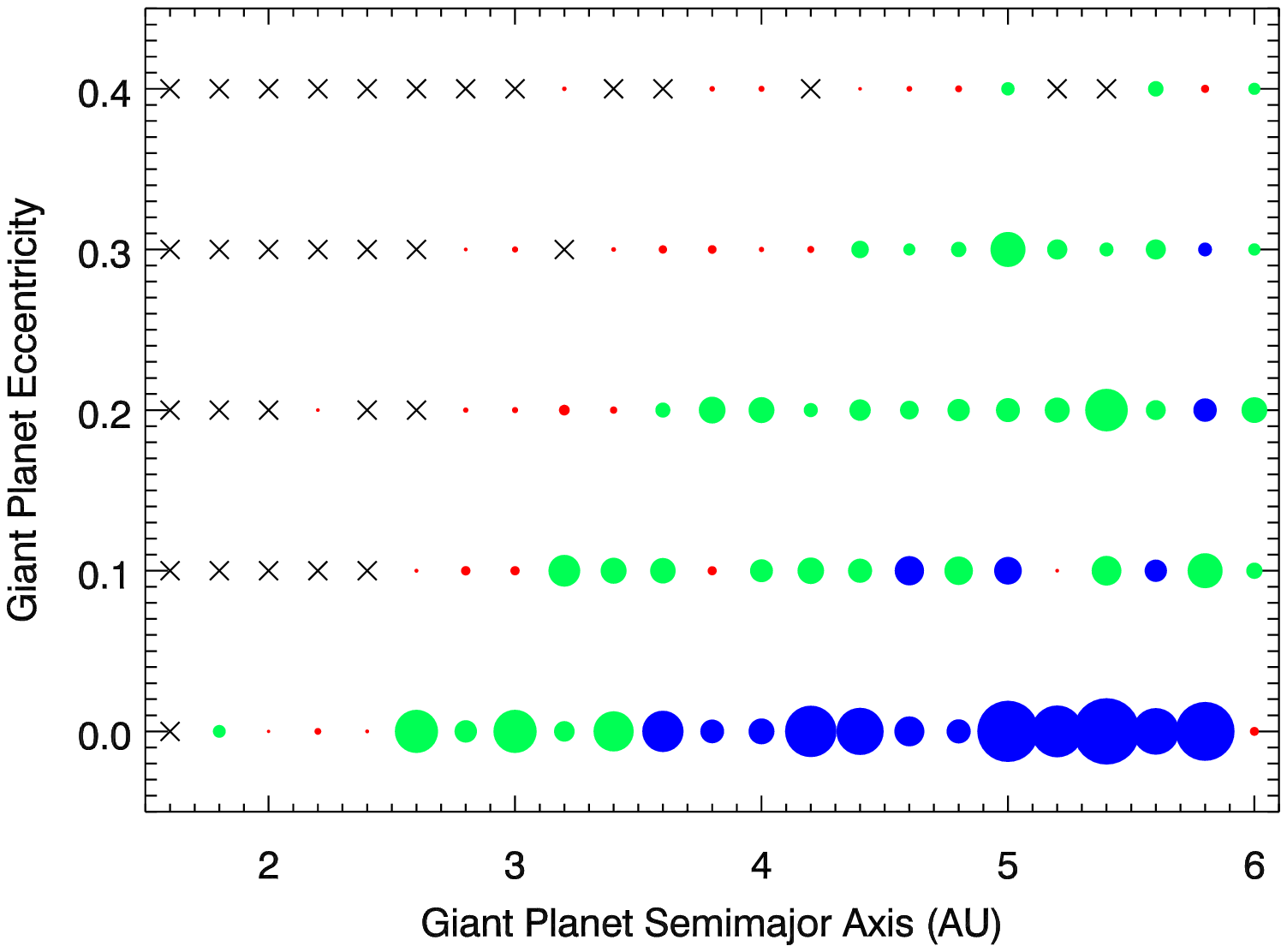}
\figcaption[f2.eps]{\label{fig:aem} \small{Regions in giant planet orbital
parameter space that permit the formation of a habitable terrestrial
planet.  Crosses indicate where no terrestrial bodies survived in the
HZ (defined as 0.8-1.5 AU).  In systems with red dots, terrestrial
bodies survived in the HZ but averaged less than 0.3 $\mearth$.  In
systems with green dots, terrestrial planets in the HZ averaged more
than 0.3 $\mearth$.  Systems with blue dots formed habitable planets
in the HZ, with masses averaging over 0.3 $\mearth$ and water contents
averaging more than $5\times10^{-4}$ (roughly the Earth's water
content).  The sizes of dots is proportional to the total surviving
mass in the HZ.}}
\medskip

These limits for Jupiter's orbital distance apply only for a single giant
planet system.  Other, more distant giant planets would induce secular
resonances (e.g., the $\nu_6$ secular resonance in the asteroid belt) and make
it more difficult for terrestrial planets to form.  Thus, if multiple giant
planets existed in a system, the limits derived above in Fig.~\ref{fig:aem}
would be lower limits -- most likely, the innermost giant planet would have to
be farther from the central star for a habitable planet to form.

At higher eccentricities the critical values of $a_J$ for habitable planet
formation are higher.  This is not surprising, as both secular and resonant
perturbations increase with eccentricity.  An eccentric giant planet ejects a
larger fraction of embryos, increases planetary eccentricities, and causes
planets to form closer to the central star than for a zero-eccentricity giant
planet (Levison \& Agnor 2003).  In addition, water-rich embryos past 2-2.5 AU
are much more likely to be ejected by an eccentric giant planet, so the
terrestrial planets' water contents are decreased (Chambers \& Cassen 2002;
RQL04).

These results are for a fixed mass protoplanetary disk.  However, observations
suggest that there exist a range of disk masses (e.g., Andre \& Montmerle
1994, Eisner \& Carpenter 2003, Andrews \& Williams 2005).  In addition, the
disk mass scales with stellar mass, with important implications for the amount
of material available for terrestrial planet formation (Raymond, Scalo \&
Meadows 2006).  The disk's density profile also affects the location of
planets that form (Raymond \etal 2005b).

\section{Application to Known Extra-Solar Systems of Giant Planets}

We adopt the two limits from Section 2: 1) habitable-mass ($>$0.3 $\mearth$)
planets can only form in the HZ if $a_J >$2.5 AU and 2) water-rich habitable
planets if $a_J >$3.5 AU.  To extend this to different stellar masses, we
assume that the ratio of orbital periods of these limits and the outer edge of
the HZ is constant.  The location of the HZ $a_{HZ}$ scales with the stellar
flux, i.e. with the stellar luminosity $L_\star$ as $L_\star^{0.5}$.  We assume
that the boundary beyond which water-rich material can exist in the disk also
scales with the stellar flux, and so do the $a_J$ limits for $>0.3\mearth$ and
water-rich planets.  For the mass-luminosity relation of low-mass stars, we
use a fit to data of Hillenbrand \& White (2004).

Figure~\ref{fig:limit} shows our derived limits on a giant planet's orbital
distance for $>0.3 \mearth$ planet formation (dashed line) and water-rich
planet formation (solid line) as a function of stellar mass.  Circles
represent the known extra solar planetary systems detected via the radial
velocity technique (with updated orbital fits and stellar data from Butler
\etal 2006 and references therein; we only considered the outermost planet in
multi-planet systems).  Since this is a slice in a many-dimensional space,
variables such as giant planet mass and eccentricity are not shown in
Fig.~\ref{fig:limit}.  Filled circles indicate giant planets with low enough
eccentricities to indeed form 0.3 $\mearth$ planets in the HZ, and filled
triangles mark the two systems that can form water-rich planets (see below).

Only nine of the 153 planetary ($\sim$6\%) systems considered fit our criteria
for being able to form habitable-mass planets.  Two of these are multi-planet
systems with an additional giant planet closer to the HZ, making it unlikely
for terrestrial bodies to survive in the HZ.  Thus, only seven of 153 (5\%)
known giant planet systems are likely to form potentially habitable planets.
Two of these are able to form water-rich planets in the HZ: 55 Cnc (as shown
by Raymond, Barnes \& Kaib 2006) and HD 89307.  The five other systems that
can form 0.3 $\mearth$ planets in the HZ are Epsilon Eridani, HD 111232, HD
114386, HD 146922, and HD 70642.  If their eccentricities are small, two
planets recently discovered by microlensing surveys would also fit both of our
$a_J$ criteria: the 5.5 $\mearth$ planet OGLE-2005-BLG-390Lb (Beaulieu \etal
2006) and the $\sim$ 13 $\mearth$ planet OGLE-2005-BLG-169 (Gould \etal 2006).

\medskip
\epsfxsize=8truecm \epsscale{0.5} \epsfbox{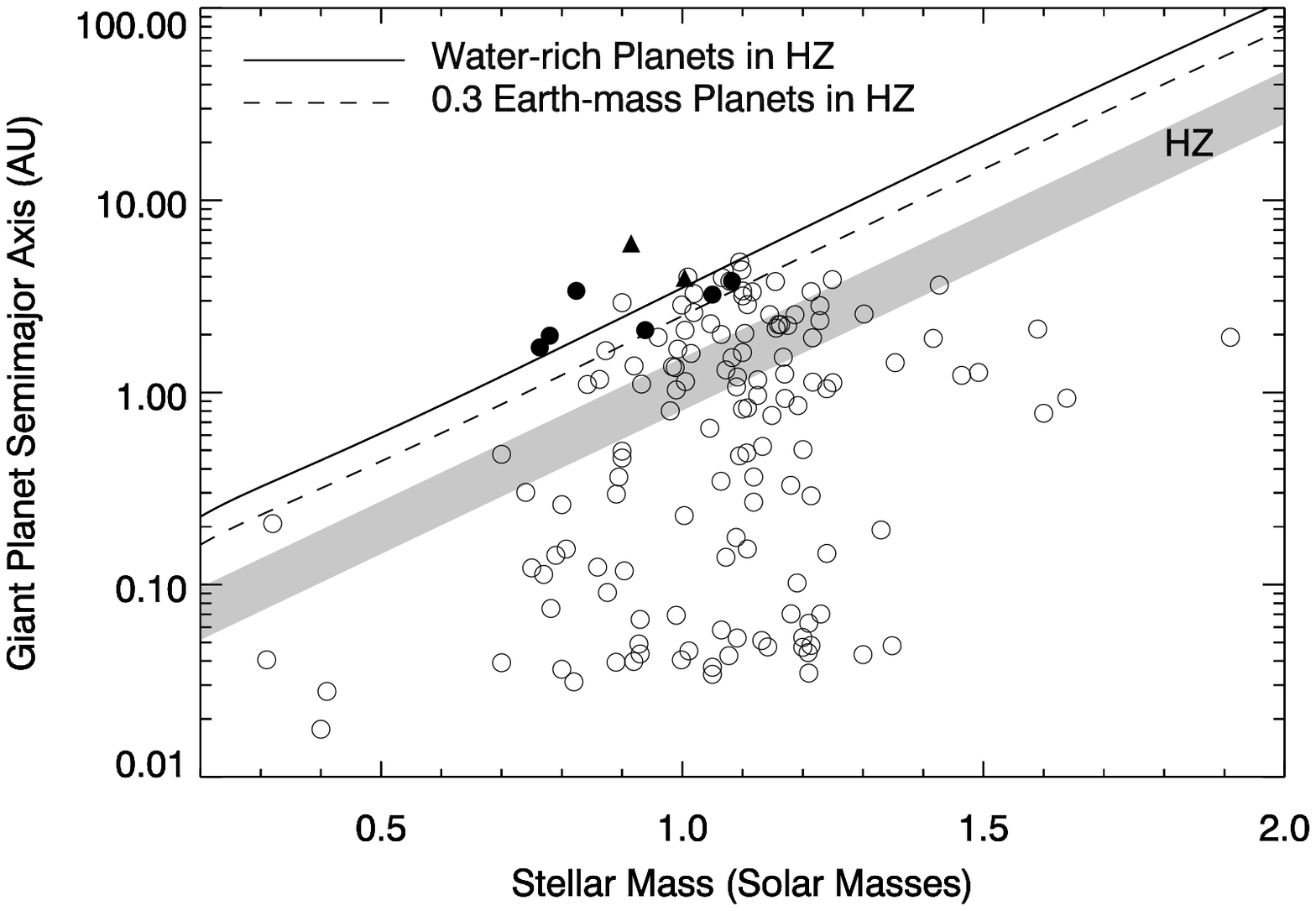}
\figcaption[f3.eps]{\label{fig:limit} \small{Limits in stellar
mass-giant planet semimajor axis space where habitable planets can form.
Above the dashed curve, 0.3 $\mearth$ planets can form in the HZ.  Above the
solid curve, water-rich 0.3 $\mearth$ planets can form in the HZ.  The open
circles indicate known giant planets.  Filled circles indicate the giant
planets which also have appropriate eccentricities for the formation of
habitable-mass planets.  Filled triangles indicate giant planets that can form
water-rich 0.3 $\mearth$ planets in the HZ.  The HZ is shaded.}}
\medskip

\section{Discussion and Conclusions}

In this {\it Letter} we have placed rough limits on where habitable
terrestrial planets can form as a function of the orbital semimajor axis $a_J$
and eccentricity $e_J$ of a Jupiter-mass giant planet.  These limits can be
used as guidelines to direct the search for extra-solar Earth-like planets.
We find that, for $>0.3 \mearth$ planets to form in the habitable zone (HZ),
$a_J$ must be larger than 2.5 AU.  If planets accrete water from volatile-rich
embryos past 2-2.5 AU, then water-rich, $>0.3 \mearth$ HZ planets can form if
$a_J>$ 3.5 AU.  If, however, embryos originating in the HZ are hydrated (Drake
\& Righter 2002), then our limit for water-laden planets is identical to our
0.3 $\mearth$ formation limit.

We have only considered Jupiter-mass giant planets.  More massive planets are
stronger perturbers and will have habitable planet limits at larger $a_J$.  In
addition, there might exist systematic links between planetary mass and
stellar mass: 1) Jupiter-mass planets may be unlikely to form around low-mass
stars (Laughlin \etal 2004), and 2) the surface density of material scales
with the stellar mass, so less material is available for building planets
around low-mass stars.  Indeed, Earth-mass planets may be rare around low-mass
stars (Raymond, Scalo \& Meadows 2006).

We have not considered giant planet migration in these simulations.  As a
giant planet forms in the outer disk and migrates inward, moving resonances
can ``push'' in front of it a pile-up of material (Fogg \& Nelson 2005;
Mandell, Raymond \& Sigurdsson 2006, in preparation).  If the giant planet
stopped migrating at 2 AU, an Earth-mass planet could potentially form just
inside the 2:1 resonance, at 1.27 AU in this case.  However, the stability of
such a planet is unclear, as it would have migrated into a disk of planetary
embryos.  Self-scattering of embryos into resonances with Jupiter is thought
to be the main cause of the depletion of the asteroid belt (Wetherill 1992).
Although such a terrestrial planet might be more massive than the embryos, it
would already lie close to strong resonances with the giant planet.  The fate
(and even the formation) of such a planet is uncertain.

Planetary systems with inner giant planets (e.g., ``hot jupiters'') may
support terrestrial planets in the HZ (Raymond \etal 2005a; Mandell \etal
2006, in preparation).  In these systems, habitable planets would be built of
rocky/icy material that survived giant planet migration through the
terresrtial region.  Raymond \etal (2005a) found that low-eccentricity giant
planets inside about 0.5 AU are able to form habitable planets.

Most of the detected giant planets are at relatively small orbital distance
and have significant orbital eccentricities.  Only seven out of 153 planetary
systems ($\sim$5\%) from Butler \etal (2006) meet our criterion for the
formation of $>0.3 \mearth$ planets, and only two meet our limit for
water-rich planets to form in the HZ.  However, if we arbitrarily assume that
habitable planets can form in systems with giant planets interior to 0.5 AU
with eccentricities less than 0.1 (roughly following from Raymond \etal
2005a), then the number of known extra solar systems that could harbor
habitable planets increases to 45 (29\%).

Many planetary systems may not harbor giant planets but still allow
terrestrial accretion to occur (Greaves \etal 2006; Raymond, Scalo \& Meadows
2006).  Thus, the abundance of potentially habitable terrestrial planets may
not be tied to the abundance of giant planets.  Extra solar Earths will be
found and characterized by upcoming missions such as {\it Kepler}, {\it
COROT}, {\it SIM}, {\it Darwin} and {\it Terrestrial Planet Finder}.  We
eagerly await these new discoveries.

\medskip

This project was inspired by Suzanne Hawley and benefited from discussions
with Avi Mandell, John Scalo and Vikki Meadows.  Many thanks to Jason Wright,
Geoff Marcy, and Paul Butler for access to their new exoplanet orbits.  Thanks
also to the anonymous referee for helpful comments.  This work was performed
by the NASA Astrobiology Institute's Virtual Planetary Laboratory Lead Team,
supported via NASA CAN-00-OSS-01.  Simulations were performed at Weber State
University and Caltech using CONDOR (www.cs.wisc.edu/condor).


\begin{thebibliography}{999}


\bibitem[]{369} Abe, Y., Ohtani, E., Okuchi, T., Righter, K., and Drake,
  M., 2000. In: Righter, K. \& Canup, R. (Eds.), Origin of the Earth and the
  Moon. University of Arizona Press, Tucson, 413-433.

\bibitem[]{373} Agnor, C.~B., Canup, R.~M., \& Levison, H.~F.\ 1999. Icarus, 142, 219. 

\bibitem[]{375} Andre, P. and Montmerle, T. 1994, ApJ, 420, 837.

\bibitem[]{377} Andrews, S. M. and Williams, J. P. 2005, ApJ, 631, 1134.

\bibitem[]{379} Barnes, R., \& Raymond, S.~N.\ 2004, \apj, 617, 569 

\bibitem[]{381} Beaulieu, J.-P., et al.\ 2006, \nat, 439, 437 

\bibitem[]{} Boss, A.~P.\ 1997, Science, 276, 1836 

\bibitem[]{} Butler, R.~P., et al. \ 2006, ApJ, in press.

\bibitem[]{383} Chambers, J. E., 1999. MNRAS, 304, 793.

\bibitem[]{385} Chambers, J.~E.\ 2001.  Icarus, 152, 205.

\bibitem[]{387} Chambers, J. E. \& Cassen, P., 2002, Meteoritics and Planetary
Science, 37, 1523

\bibitem[]{390} Chambers, J.~E.\ 2004, Earth  and Planetary Science
Letters, 223, 241 

\bibitem[]{393} Endl, M., Cochran, W. D., Kurster, M., Paulson, D. B.,
Wittenmyer, R. A., MacQueen, P. J., and Tull, R. G. 2006, ApJ,
submitted.

\bibitem[]{397} Eisner, J. A. and Carpenter, J. M.  2003, ApJ, 598, 1341.

\bibitem[]{399} Fischer, D. A. and Valenti, J. 2005. ApJ 622, 1102

\bibitem[]{401} Fogg, M.~J., \& Nelson, R.~P.\ 2005, \aap, 441, 791 

\bibitem[]{} Gould, A., et al.\ 2006, ApJL, submitted, astro-ph/0603276

\bibitem[]{403} Greaves, J.~S., Fischer, D.~A., \& Wyatt, M.~C.\ 2006, \mnras, 366, 283 

\bibitem[]{405} Hayashi, C. 1981.  Prog. Theor. Phys. Suppl., 70, 35.

\bibitem[]{407} Haisch, K.~E., Lada, E.~A., \& Lada, C.~J.\ 2001, \apjl, 553, L153

\bibitem[]{409} Hillenbrand, L.~A., \& White, R.~J.\ 2004, \apj, 604, 741 

\bibitem[]{411} Jones, B.~W., Sleep, P.~N., \& Chambers, J.~E.\ 2001, \aap, 366, 254 

\bibitem[]{413} Kasting, J. F., Whitmire, D. P., and Reynolds, R. T.,
  1993. Icarus 101, 108.

\bibitem[]{416} Kokubo, E., \& Ida, S.\ 1998, Icarus, 131, 171 

%\bibitem[]{418} Kokubo, E., Kominami, J., \& Ida, S., 2006.  ApJ, in press.

\bibitem[]{420} Laughlin, G., Bodenheimer, P., \& Adams, F.~C.\ 2004,
\apjl, 612, L73 

\bibitem[]{423} Levison, H. F. \& Agnor, C., 2003, \aj, 125, 2692.

\bibitem[]{425} Lineweaver, C. H. and Grether, D. 2003, ApJ 598, 1350

\bibitem[]{427} Mandell, A., Raymond, S.~N., \& Sigurdsson, S. 2006, In preparation.

\bibitem[]{429} Menou, K., \& Tabachnik, S.\ 2003, \apj, 583, 47

\bibitem[]{431} Morbidelli, A., Chambers, J., Lunine, J. I., Petit, J. M.,
Robert, F., Valsecchi, G. B., and Cyr, K. E., 2000. Meteoritics and
Planetary Science 35, 1309.

%\bibitem[]{} Petit, J.-M., Morbidelli, A., \& Chambers, J.\ 2001, Icarus, 153, 338

\bibitem[]{} Pollack, J.~B., Hubickyj, O., Bodenheimer, P., Lissauer, J.~J.,
Podolak, M., \& Greenzweig, Y.\ 1996, Icarus, 124, 62

\bibitem[]{435} Quintana, E.~V., Lissauer, J.~J., Chambers, J.~E., \& Duncan, M.~J.\
2002, \apj, 576, 982

\bibitem[]{438} Raymond, S.~N., Quinn, T., \& Lunine, J. I., 2004. (RQL04) Icarus,
168, 1.

\bibitem[]{441}Raymond, S.~N., Quinn,~T., \& Lunine,~J.~I. 2005a. Icarus,
177, 256.

\bibitem[]{444}Raymond, S.~N., Quinn,~T., \& Lunine,~J.~I. 2005b. \apj, 632, 670.

\bibitem[]{446}Raymond, S.~N., Quinn,~T., \& Lunine,~J.~I. 2006a. Icarus,
in press, astro-ph/0510284.

\bibitem[]{449}Raymond, S.~N., Quinn,~T., \& Lunine,~J.~I. 2006b.
Astrobiology, submitted, astro-ph/0510285.

%\bibitem[]{} Raymond, S.~N., \& Barnes, R.\ 2005, \apj, 619, 549

\bibitem[]{452}Raymond, S.~N., Barnes, R., \& Kaib, N.~A. 2006.  \apj, in press, astro-ph/0404212

\bibitem[]{454} Raymond, S.~N., Scalo, J., \& Meadows, V.~S. 2006.  ApJL,
submitted.

\bibitem[]{457} Regenauer-Lieb, K., Yuen, D. A., \& Branlund, J. 2001, Science
294, 578.

\bibitem[]{460}Rivera, E.~J. \& Lissuaer, J.~J. 2001, \apj, 558, 392

\bibitem[]{462} Tabachnik, S. and Tremaine, S. 2002, MNRAS 335, 151.

\bibitem[]{464} Th{\'e}bault, P., Marzari, F., \& Scholl, H.\ 2002, \aap, 384, 594 

\bibitem[]{466} Th{\'e}bault, P., Marzari, F., Scholl, H., Turrini, D., \&
Barbieri, M.\ 2004, \aap, 427, 1097 

\bibitem[]{469} Wetherill, G. W. 1992, Icarus, 100, 307

\bibitem[]{471} Williams, D.~M., Kasting, J.~F., \& Wade, R.~A.\ 1997, Nature, 385, 234. 

\end{thebibliography}
\end{document}